\documentclass[a4paper,fleqn,usenatbib]{mnras}
\usepackage[T1]{fontenc}
\usepackage{ae,aecompl}

\usepackage{graphicx}	
\usepackage{amsmath}	
\usepackage{amssymb}	

\newcommand{\dw}{LUH\,16}

\title[Mass ratio of \dw]{Mass ratio of the 2\,pc binary brown dwarf   \dw\ and limits on planetary companions from astrometry\thanks{Based on data obtained from the ESO Science Archive Facility under programme IDs 291.C-5004 and 593.C-0314.}} 

\author[Sahlmann \& Lazorenko]
{J. Sahlmann$^{1}$\thanks{E-mail: Johannes.Sahlmann@esa.int}\thanks{ESA Research Fellow}
and P. F. Lazorenko$^{2}$ \\
$^{1}$European Space Agency, European Space Astronomy Centre, P.O. Box 78, Villanueva de la Ca\~nada, 28691 Madrid, Spain\\
$^{2}$Main Astronomical Observatory, National Academy of Sciences of the Ukraine, Zabolotnogo 27, 03680 Kyiv, Ukraine}

\date{Accepted 2015 mmm dd. Received 2015 mmm dd; in original form 2015 June 12}
\pubyear{2015}

\begin{document}
\label{firstpage}
\pagerange{\pageref{firstpage}--\pageref{lastpage}}
\maketitle

\begin{abstract}
We analyse FORS2/VLT $I$-band imaging data to monitor the motions of both components in the {nearest} known binary brown dwarf WISE J104915.57-531906.1AB (\dw) over one year. The astrometry is dominated by parallax and proper motion, but with a precision of $\sim$0.2 milli-arcsecond per epoch we accurately measure the relative position change caused by the orbital motion of the pair. This allows us to directly {measure} a mass ratio of $q=0.78\pm0.10$ for this system. We also search for the signature of a planetary-mass companion around either of the A and B component and exclude at 3-$\sigma$ the presence of planets with masses larger than $2\,M_\mathrm{Jup}$ and orbital periods of 20--300 d. We update the parallax of \dw\ to $500.51\pm0.11$ mas, i.e.\ just within 2 pc. This study yields the first direct constraint on the mass ratio of \dw\ and shows that the system does not harbour any close-in giant planets. 
\end{abstract}


\section{Introduction}
The  binary brown dwarf \href{http://simbad.u-strasbg.fr/simbad/sim-basic?Ident=WISE+J104915.57-531906.1\&submit=SIMBAD+search}{WISE J104915.57-531906.1} (**LUH\,16), hereafter \dw\ \citep{Luhman:2013aa} represents a unique opportunity to study the properties of substellar objects because of its proximity of 2 pc to  the Sun. The system consists of a primary with spectral type L7.5 (component A) and a T0.5 secondary (component B) \citep{Burgasser:2013aa}. Both show strong Li I absorption in their optical spectra \citep{Faherty:2014aa}, which confirms that they are the {nearest} brown dwarfs yet discovered. The system's orbital period was estimated at 25-30 years \citep{Luhman:2013aa,Mamajek:2013aa} and measuring the orbital motion offers the opportunity of constraining the component masses, hence determining yet unknown fundamental parameters of the system. \dw\ is also an excellent target to search for orbiting planets \citep{Gillon:2013aa,Boffin:2014aa}, allowing us to investigate planet formation around brown dwarfs. 
\section{Data reduction}
\begin{figure} 
\centering
\includegraphics[width=0.7\linewidth]{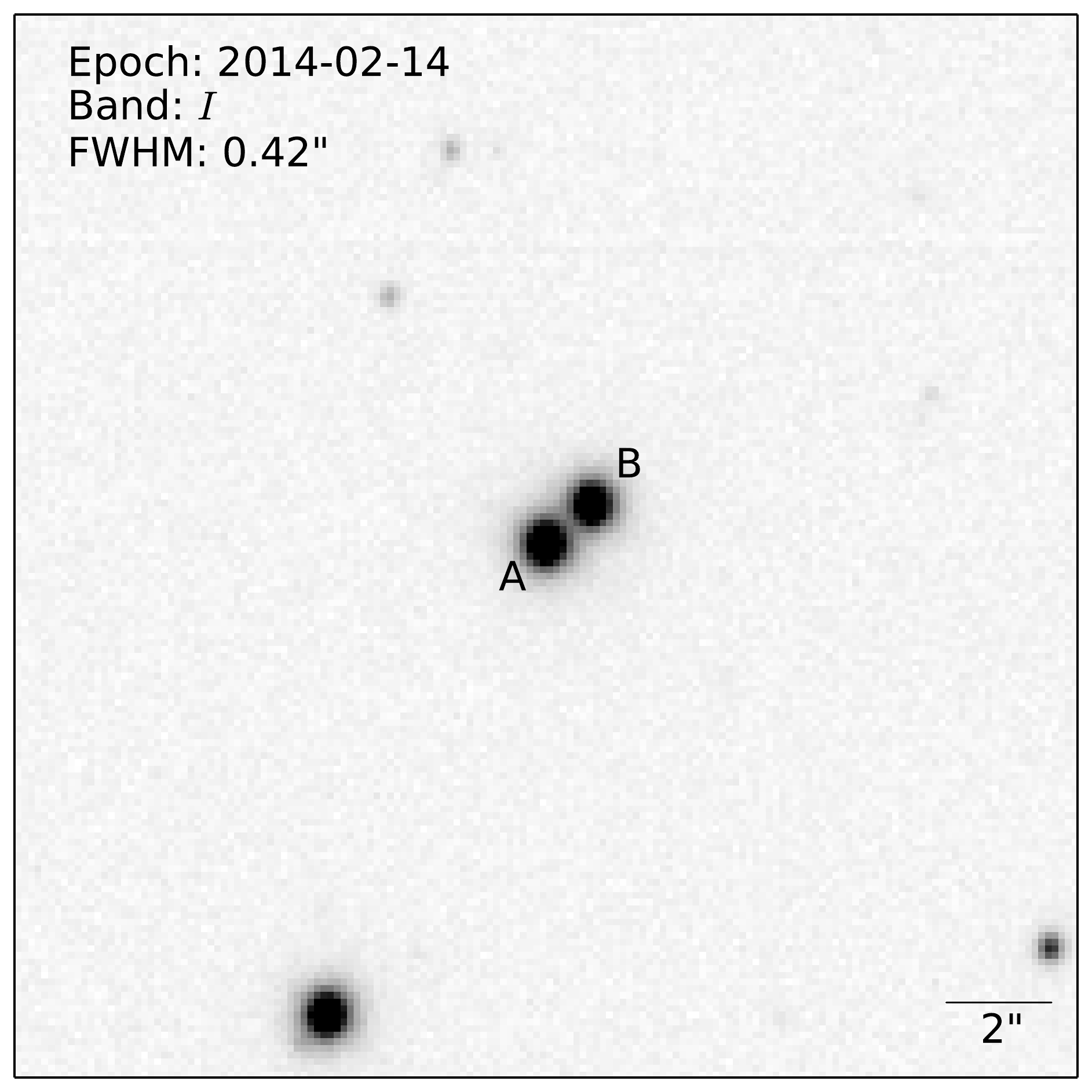}
\caption{FORS2 image of \dw\ in very good seeing conditions where the A and B components are well resolved. The image size is $\sim$$20\arcsec\! \times 20\arcsec$, i.e.\ only a small fraction of the $4\arcmin\times4\arcmin$ field of view, with north up and east left. The star to the south-east is USNO-B1.0 0366-0244607 ($I=15.7$).}
\label{fig:image}
\end{figure}

We analysed FORS2 images of \dw\ taken between April 2013 and May 2014 that were retrieved from the ESO archive (Programmes 291.C-5004 and 593.C-0314; PI: Boffin). The instrument setup and observation strategy used in these programmes ($I$-Bessel filter, several dithered frames per epoch, target position on CCD chip1, constraints on airmass and atmospheric conditions) are very similar to those of our exoplanet search survey \citep{Sahlmann:2014aa}  and we reduced the data with our methods developed for that purpose \citep{Lazorenko:2009ph,Lazorenko:2014aa}.  Table \ref{tab:obs} summarises the data we used and lists the epoch number, the mean date of the epoch exposures, the average airmass and the average FWHM measured for star images. There are 22 epochs spanning 399 d, and every epoch consists of 16 to 42 usable individual exposures ($N_\mathrm{f}$) taken over $\Delta t=0.5$ h on average, resulting in a total of 583 exposures. Figure \ref{fig:image} shows an example image.

We used 585 reference stars located within a radius of 2\farcm2 of \dw\ to measure the motion of the binary relative to the background field. {They span a magnitude range of $I\sim15-21$, where 41 stars are within $\pm$1 mag of \dw A.} The components A and B were resolved in the images and we obtained their individual astrometry with an average per-epoch precision of 0.15 and 0.17 milli-arcsecond (mas), respectively.

\begin{table}
\caption{FORS2 data used in the astrometric analysis.}
\centering
\begin{tabular}{rcrrrrcc}
\hline
No & Mean date & $N_\mathrm{f}$ & $\Delta t$ & Air- & FWHM \\
 & (UT) &  & (h) & mass& (\arcsec) \\
\hline
1 & {2013-04-14T23:52:16} & 16 & 0.26 & 1.25 & 0.69 \\
2 & {2013-04-20T01:28:07} & 21 & 0.31 & 1.14 & 0.66 \\
3 & {2013-04-26T00:58:26} & 21 & 0.26 & 1.14 & 0.56 \\
4 & {2013-05-06T00:46:31} & 17 & 0.36 & 1.14 & 0.83 \\
5 & {2013-05-12T02:05:37} & 21 & 0.27 & 1.23 & 0.81 \\
6 & {2013-05-20T23:26:08} & 30 & 0.76 & 1.14 & 0.78 \\
7 & {2013-05-26T00:45:33} & 19 & 0.26 & 1.20 & 0.58 \\
8 & {2013-05-31T23:48:51} & 21 & 0.26 & 1.16 & 0.68 \\
9 & {2013-06-04T23:57:18} & 21 & 0.27 & 1.19 & 0.66 \\
10 & {2013-06-09T23:18:32} & 41 & 0.58 & 1.17 & 0.69 \\
11 & {2013-06-13T23:56:37} & 34 & 0.53 & 1.24 & 0.70 \\
12 & {2013-06-16T23:23:53} & 20 & 0.32 & 1.20 & 0.71 \\
13 & {2013-06-22T23:13:02} & 24 & 0.43 & 1.22 & 0.85 \\
14 & {2014-02-05T06:00:02} & 26 & 0.67 & 1.15 & 0.74 \\
15 & {2014-02-14T04:32:37} & 42 & 0.64 & 1.19 & 0.52 \\
16 & {2014-03-10T03:31:31} & 18 & 0.61 & 1.16 & 0.68 \\
17 & {2014-03-19T04:21:07} & 34 & 0.66 & 1.15 & 0.77 \\
18 & {2014-03-30T02:57:14} & 32 & 0.67 & 1.14 & 0.75 \\
19 & {2014-04-10T02:45:09} & 32 & 0.67 & 1.15 & 0.68 \\
20 & {2014-04-25T23:58:25} & 30 & 0.67 & 1.18 & 0.66 \\
21 & {2014-05-05T23:44:34} & 29 & 0.65 & 1.16 & 0.79 \\
22 & {2014-05-18T23:35:28} & 34 & 0.67 & 1.14 & 0.75 \\
\hline
\end{tabular}
\label{tab:obs}
\end{table}

\vspace{-5mm}
\section{Mass ratio measurement}
The standard model for FORS2 astrometry has seven free parameters ($\Delta\alpha^{\star}_0, \Delta\delta_0, \mu_{\alpha^\star}, \mu_\delta, \varpi$, $\rho$, and $d$) and reproduces the astrometric measurements of a target $\alpha^{\star}_m$\footnote{We use the notation $\alpha^{\star} = \alpha \cos{\delta}$ throughout the text.} and $\delta_m$ in RA and Dec, respectively, in frame $m$ at time $t_m$ relative to the reference frame of background stars \citep{Lazorenko:2011lr,Sahlmann:2014aa}:
\begin{equation}\label{eq:axmodel}
\begin{array}{ll@{\hspace{2mm}}l}
\!\alpha^{\star}_{m} =\!\!\!\!\!& \Delta \alpha^{\star}_0 + \mu_{\alpha^\star} \, t_m + \varpi \, \Pi_{\alpha,m}\!\!\! &- \rho\, f_{1,x,m} -  d \,f_{2,x,m} \\
\delta_{m} =\!\!\!\!\!&{\Delta \delta_0 + \,\mu_\delta      \,  \;                      t_m \;+ \varpi \, \Pi_{\delta,m}} \!\!\! &{+ \rho \,f_{1,y,m}} +  d \,f_{2,y,m},
\end{array}
\end{equation}
where $\Delta\alpha^{\star}_0, \Delta\delta_0$ are coordinate offsets, $\mu_{\alpha^\star}, \mu_\delta$ are relative proper motions, and the parallactic motion is expressed as the product of relative parallax $\varpi$ and the parallax factors $\Pi_\alpha, \Pi_\delta$. The parameters $\rho$ and $d$ model differential chromatic refraction (DCR) and their coefficients $f_{1,2}$ are functions of  zenith angle, temperature, and pressure \citep{Lazorenko:2006qf, Sahlmann:2013ab}. 

{The epoch-averaged astrometry of \dw A and B is given in Table \ref{tab:epastr}. Transformation to the ICRF was performed
using 118 reference stars that are included in the USNO-B catalogue \citet{Monet:2003rt}. The residual dispersion of this mapping leads to an uncertainty of 80 mas in the absolute coordinates of \dw. However, all our results were obtained by fitting the 583 single-frame astrometric measurements, which are available upon request.}

\subsection{Individual astrometry}\label{sec:ind}
We first modelled the individual astrometry $\alpha^{\star}_\mathrm{A}$,$\delta_\mathrm{A}$ and $\alpha^{\star}_\mathrm{B}$,$\delta_\mathrm{B}$ of \dw\ A and B separately using Eq. (\ref{eq:axmodel}). We abandoned this approach because the resulting proper motions differed significantly between A and B by $-212.1 \pm 0.2$ and $322.1 \pm 0.2$ mas/yr in $\mu_{\alpha^\star}$ and $\mu_\delta$, respectively, which is unphysical for a gravitationally bound system. In addition, the epoch residuals for A (0.9 mas r.m.s.) and B (1.1 mas r.m.s.) were anti-correlated and much larger than our measurement uncertainties, indicating that the motions of A and B are not independent and should be modelled globally. 

\subsection{Orbital motion}
\begin{figure} 
\centering
\includegraphics[width=0.8\linewidth]{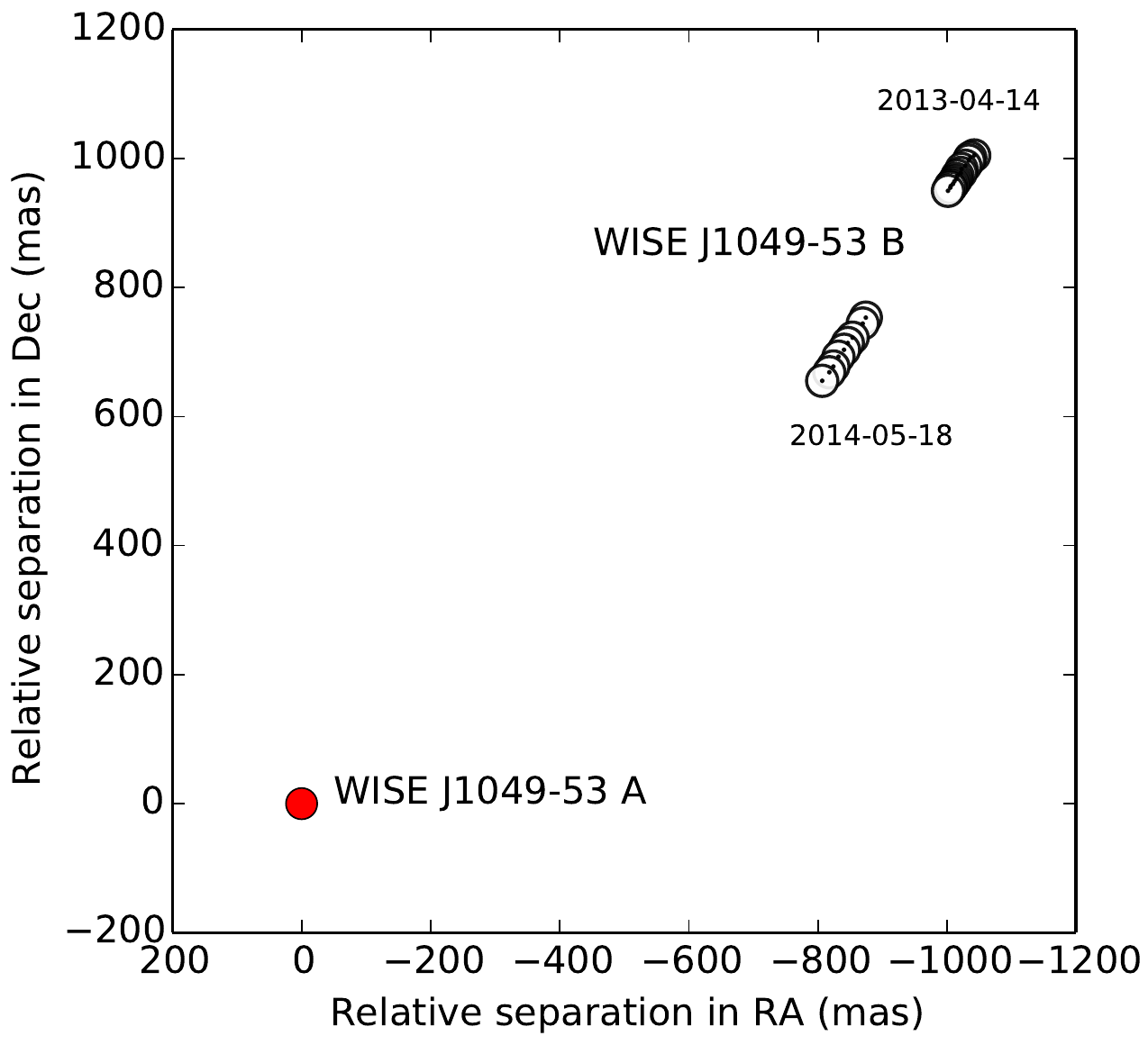}
\caption{Position change of \dw\ B (open circles) relative to \dw A (filled red circle) over the course of the observations, corresponding to a section of the system's orbit, which has to be highly inclined or very eccentric. Uncertainties are much smaller than the symbol sizes.}
\label{gif:relsep}
\end{figure}

Figure \ref{gif:relsep} shows the position of component B relative to A obtained as $\alpha^{\star}_\mathrm{B}-\alpha^{\star}_\mathrm{A}$  and $\delta_\mathrm{B}-\delta_\mathrm{A}$ for every observation epoch. Over the course of the observations, the relative separation decreases by 0\farcs41 whereas the position angle decreased only slightly by 4.9\degr\ from its initial value of 314\degr. We thus detect the effect of the orbital motion of the system, with an amplitude that agrees with the prediction of \cite{Burgasser:2013aa} and that is consistent with the $\sim$1\farcs5 separation measurement of \citet{Luhman:2013aa}. {The direction of the relative motion} indicates a highly inclined or very eccentric orbit.  

\subsection{Barycentre motion and mass ratio}
In a binary system, the astrometric orbital motions of the primary and the secondary relative to the barycentre are determined by the orbital period $P$, inclination relative to the sky plane $i$, eccentricity $e$, argument $\omega$ and time $T_0$ of periastron, ascending node $\Omega$, and the component masses $m_A$ and $m_B$ \citep[e.g.][]{Hilditch:2001kx}. The latter are related to the barycentric semimajor axes $\bar a_A$ and $\bar a_B$ via $\bar a_A/\bar a_B = m_B/m_A=q$, where $q$ is the mass ratio. Because the orbital parameters of A and B only differ by 180\degr\ in the argument of periastron, the position $\alpha^{\star}_\gamma$,$\delta_\gamma$ of the barycentre at any time is tied to the positions of the individual components:
\begin{equation}\label{eq:barycentre}
\begin{array}{ll}
\!\alpha^{\star}_{\gamma} = &\frac{1}{1+q} \left( \alpha^{\star}_\mathrm{A} + q\, \alpha^{\star}_\mathrm{B}\right) \\[2pt]
\delta_{\gamma} = &\frac{1}{1+q} \left( \,\delta_\mathrm{A} + q \,\,\delta_\mathrm{B}\right). \\
\end{array}
\end{equation}
We thus computed the barycentre positions of \dw\ for a range of mass ratios according to Eq. (\ref{eq:barycentre}) and fitted them with the model Eq. (\ref{eq:axmodel}). The covariance matrices associated with the measurements of A and B were accounted for accordingly and a linear least-squares fit \citep[cf.][]{Sahlmann:2014aa} was used to obtain the model parameters and their uncertainties for every value of $q$. 

\begin{figure} 
\centering
\includegraphics[width=\linewidth]{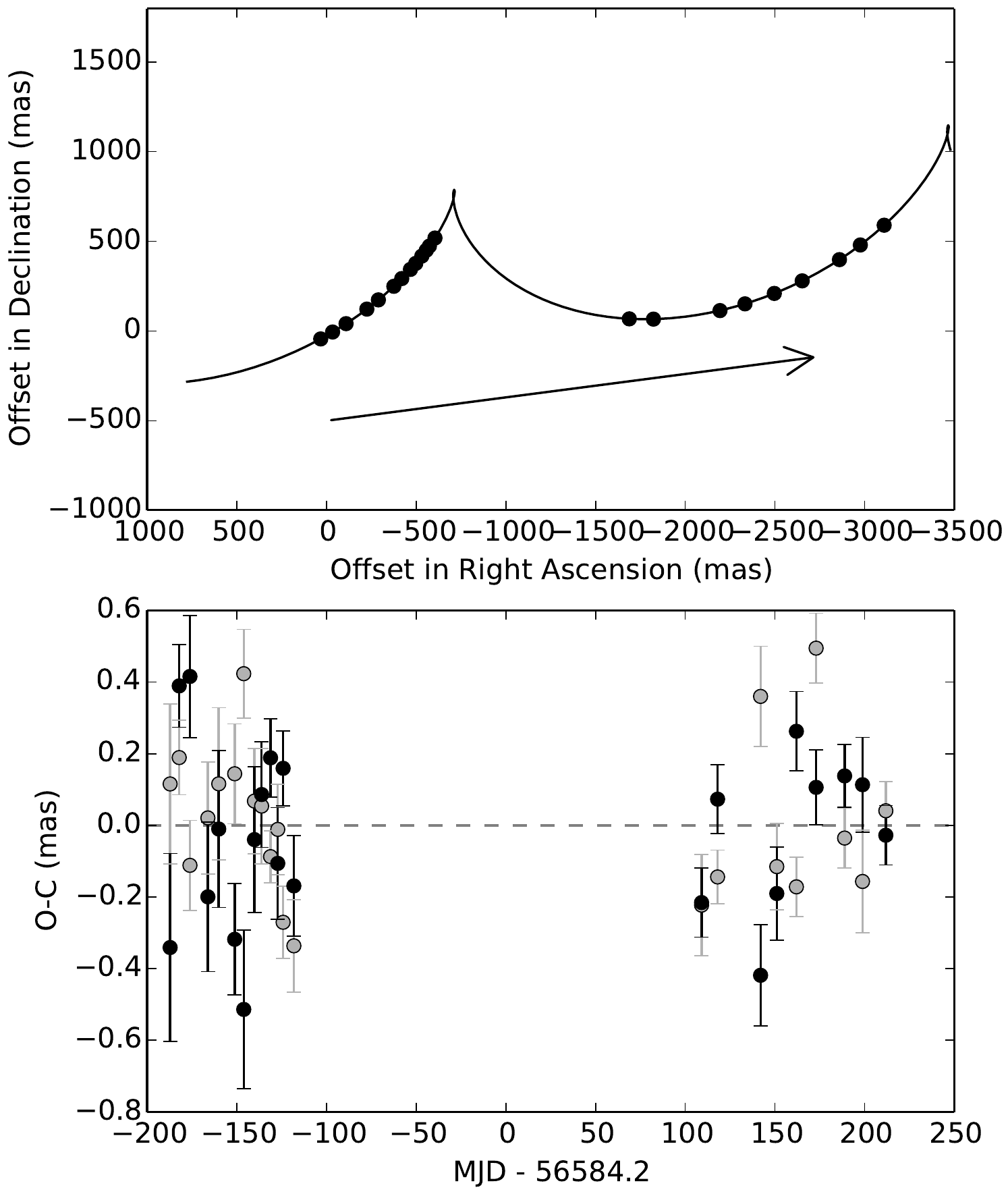}
\caption{\emph{Top:} The sky-projected barycentre motion of \dw\ measured with FORS2. Individual epoch measurements are shown with black circles and the model corresponding to the best-fit mass ratio of $q=0.78$ is shown by the curve. The arrow indicates the proper motion per year. \emph{Bottom:} Epoch residuals in RA (grey symbols) and Dec (black symbols) of the 7-parameter fit.}\label{fig:baryfit}
\end{figure}

Figure \ref{fig:baryfit} shows the result for the mass ratio producing the smallest epoch $\chi^2$. The residual RMS is 0.23 mas, i.e.\ much smaller than for the individual fits discussed in Sect. \ref{sec:ind}. However, it is still larger than the average epoch precision of 0.11 mas. We attribute this to the decreasing relative separation of the pair, which in the second season approaches the median FWHM value of 0\farcs7. As a consequence, the images of A and B overlap more and more, which introduces errors of a few milli-pixel in our photocentre estimation. 

\begin{figure} 
\centering
\includegraphics[width=0.95\linewidth]{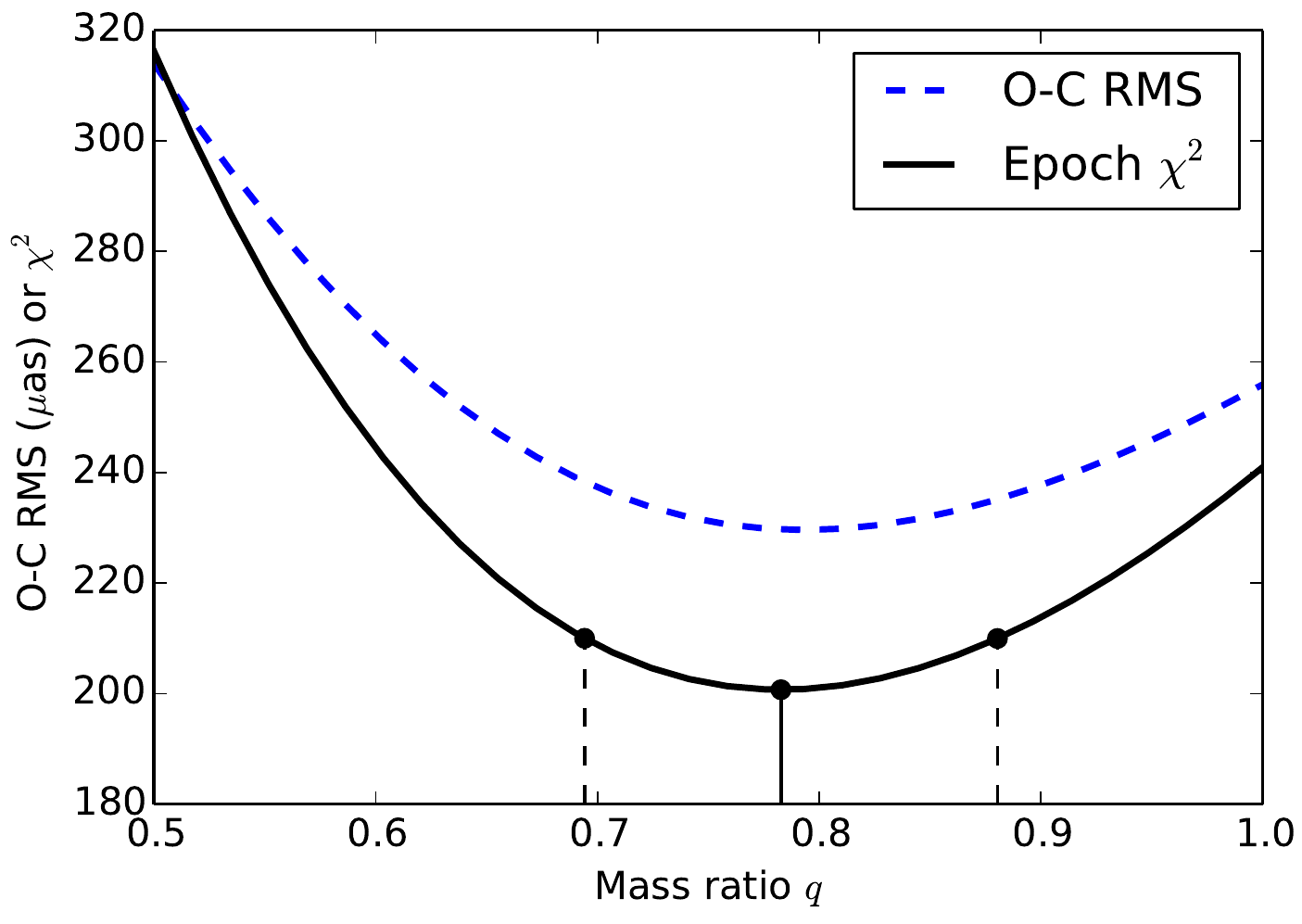}
\caption{Epoch $\chi^2$ (solid curve) and residual r.m.s. (dashed curve) in micro-arcsecond ($\mu$as) of the barycentre astrometry fit as a function of mass ratio. The solid vertical line indicates the minimum $\chi^2$ value and the dashed lines indicate the 1-$\sigma$ confidence intervals determined by $\Delta \chi^2=9.3$.}
\label{fig:chi}
\end{figure}

To determine the mass ratio of \dw, we inspected the epoch $\chi^2$ metric shown in Fig. \ref{fig:chi}, which has a parabolic shape with a well-defined minimum. The $\chi^2$ is minimum for $q=0.78 ^{+0.10}_{-0.09}$, where we derived the 1-$\sigma$ confidence interval from the $q$-values where the curve reaches a $\Delta \chi^2=+9.3039$, appropriate for a fit with 8 free parameters. Table \ref{tab:res} shows the best-fit parameters at optimal $\chi^2$ and their systematic variation with $q$ within its 1-$\sigma$ interval.

\begin{figure} 
\centering
\includegraphics[width=\linewidth]{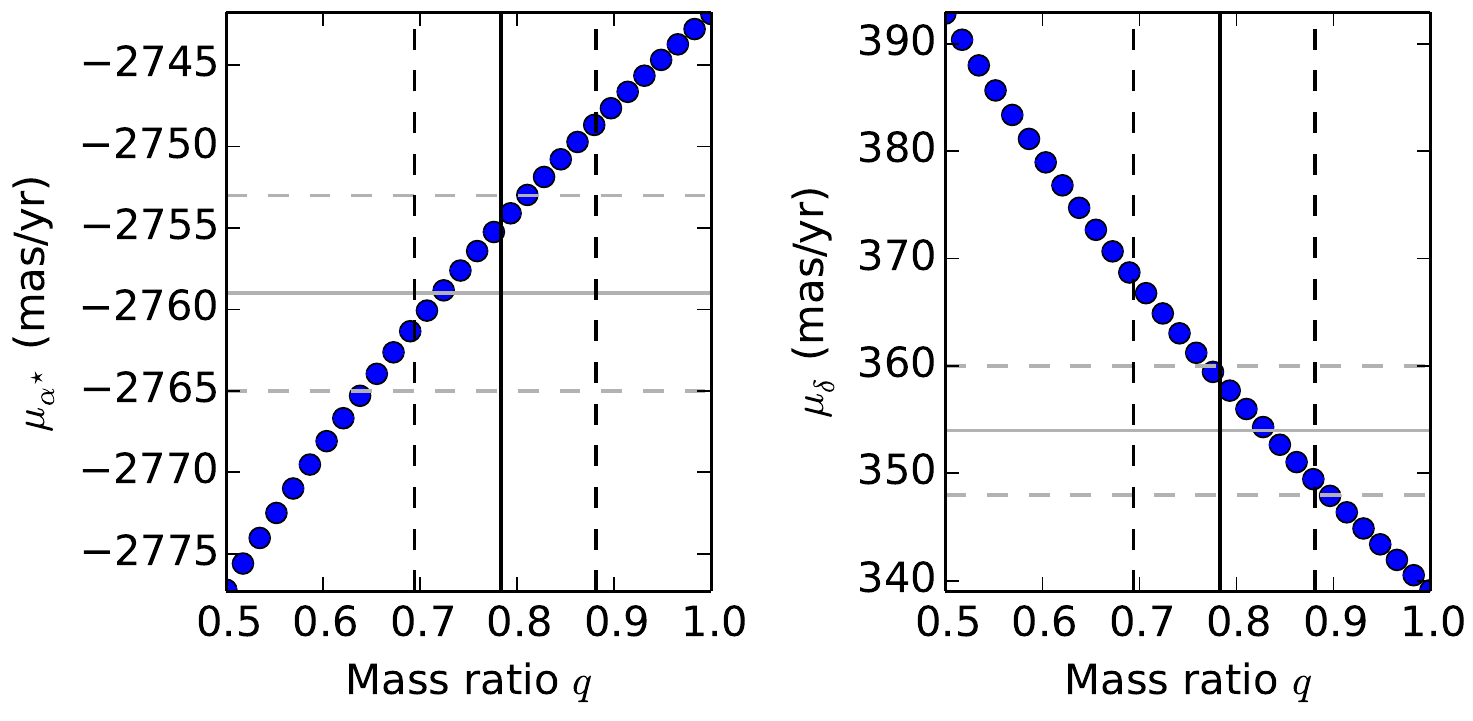}
\caption{Correlations between $q$ and proper motion in RA (left) and Dec (right). Blue symbols with uncertainties show the results of the linear fit. Horizontal lines indicate the values and uncertainties given by \citet{Luhman:2013aa} and vertical lines indicate the adopted $q$ value.} \label{fig:qcorrel}
\end{figure}

In addition to the minimum $\chi^2$ criterion, we also inspected the correlations between the mass ratio value and the derived proper motions. Because proper motions have been measured with high relative precision from data spanning 33 years \citep{Luhman:2013aa}, we can use them to determine which $q$-values are meaningful\footnote{We neglected the difference between absolute and relative proper motions, because we estimated that it is smaller than 1 mas/yr for our FORS2 measurements.}. As shown in Fig. \ref{fig:qcorrel}, there is good agreement between the $q$ values determined via $\chi^2$ and the range of values that are compatible with the proper motions of \cite{Luhman:2013aa}.

\begin{table}
\caption{Astrometric parameters of \dw\ determined solely by FORS2 data. The last two columns show the variation that corresponds to $q = q_0 \pm \sigma_q$.}
\centering
\begin{tabular}{ccrrr}
\hline
Param & Unit & Formal value & \multicolumn{2}{c}{Interval}\\
\hline
$q$ & & 0.78 & $0.69$ & $0.88$ \\
$\Delta\alpha^{\star}_0$ & mas & $-1054.72\pm0.05$ & ${+2.85}$ & ${-2.83}$ \\
$\Delta\delta_0$ & mas & $389.24\pm0.08$ & ${+4.53}$ & ${-4.49}$ \\
$\varpi_{\mathrm{rel}}$ & mas & $500.23\pm0.06$ & ${+0.11}$ & ${-0.11}$ \\
$\mu_{\alpha^\star}$ & mas/yr & $-2754.77\pm0.09$ & ${+6.25}$ & ${-6.20}$ \\
$\mu_\delta$ & mas/yr & $358.72\pm0.09$ & ${+9.49}$ & ${-9.41}$ \\
$\rho_{\mathrm{eff}}$ & mas & $35.23\pm1.90$ & ${+0.13}$ & ${-0.13}$ \\
$d_{\mathrm{eff}}$ & mas & $-48.71\pm1.56$ & ${+0.18}$ & ${-0.17}$ \\
\hline
\end{tabular}
\label{tab:res}
\end{table}

\subsection{Updated parallax}
The analysis above also allows us to determine the parallax of \dw\ with high precision. To convert the relative parallax in Table \ref{tab:res} to absolute, we used the methods described in \cite{Sahlmann:2014aa,Sahlmann:2013ab} and the Galaxy model of \cite{Robin:2003fk} to estimate a parallax correction of $0.28 \pm 0.01$ mas, using the measured parallaxes of 175 reference stars. Taking into account the mass ratio uncertainty, we thus update the absolute parallax of \dw\ to $500.51\pm0.11$ mas, which agrees with the estimate of \citet{Luhman:2013aa} and corresponds to a distance of $1.9980 \pm 0.0004$ pc. 

\subsection{Prospect for measuring individual masses}
\begin{figure} 
\centering
\includegraphics[width=0.7\linewidth]{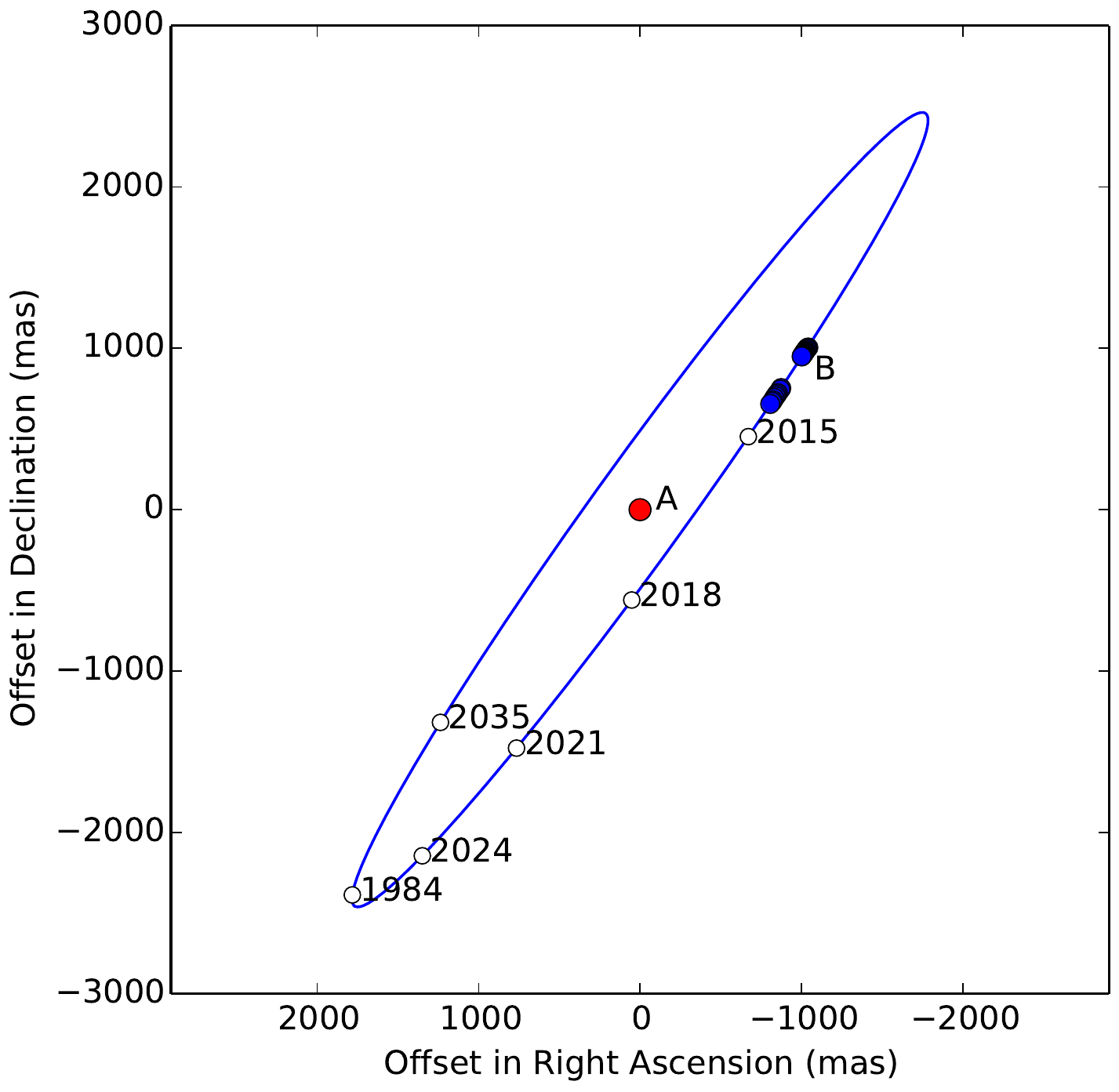}
\caption{Possible relative orbit of \dw\ with a period of 45 y, derived under the assumption that is is circular.} 
\label{fig:relOrbit}
\end{figure}

Whereas currently the estimation of individual masses relies on theoretical models, it will be possible to determine them accurately by mapping the relative orbit over about half of a revolution. The physical relative semimajor axis is proportional to $M_A+M_B$, which in combination with our mass ratio measurement yields direct individual masses. We attempted to constrain the relative orbit using the small arc measured with FORS2, by considering only two extreme configurations: a face-on eccentric orbit and an inclined circular orbit. A face-on orbit is highly unlikely because it requires extreme eccentricities ($>0.95$) and very short orbital periods ($<8$ years) to come close to reproduce the measurements. On the other hand, a circular orbit provides an acceptable fit to the data (residual epoch r.m.s. of $\sim$1.6 mas) for an inclination of $95\degr$ and an orbital period of 45 years, which is slightly longer than previous period estimates \citep{Mamajek:2013aa, Burgasser:2013aa}. The configuration shown in Fig. \ref{fig:relOrbit} is the best fit to the data with 4 free parameters ($M_A=0.06\,M_{\odot}$, $q=0.78$, $e=0$, $\omega=180\degr$ were fixed) and is compatible with the elongation of the ESO optical image in 1984 discussed by \citet{Mamajek:2013aa}. A highly inclined orbit is the most likely to be observed for a ensemble of randomly oriented orbits, however, this simplified and tentative orbit will have to be revised with future measurements. If the 45 year period is accurate, we estimate that reliable individual masses can be obtained around the year 2030. 

\section{Limits on a tertiary component}
The small residuals of the barycentre fit presented in the previous section assert that our model represents the data well. We can thus use the data to derive limits on a possible third body in the system, which would introduce an additional astrometric signal. We first searched for significant signals in the residual periodogram, finding none. At a fixed distance, the astrometric signature of an unseen companion depends on its mass, the orbital period, and the geometric configuration of the system. Accounting for the actual FORS2 sampling and the accuracy of our measurements, we then derived companion detection limits as a function of period and mass.  

We followed the procedure of \cite{Sahlmann:2014aa}, which consists in simulating 5000 circular orbits for each point in a grid of 100 periods (20--1000 d) and 60 companion masses (0.1--5\,$M_J$), i.e. 30 million pseudo-orbits, with randomly assigned parameters $\Omega$, $T_0$, $i$, the latter with a probability density function of $\frac{1}{2}\sin i$. For every orbit, the astrometric signal is computed at the sampling times and added to the observed astrometry, followed by the linear fit using Eq. \ref{eq:axmodel}. A companion of given mass and period is considered detected if 99.7\% of the corresponding pseudo-orbits have a residual r.m.s. larger than the one observed for the target. This procedure yields 3-$\sigma$ mass detection limits as a function of period.  

We run these simulations for both components of \dw, where an additional step is required because the reference residual r.m.s. was obtained for the barycentre.  A photocentre shift of component A and B will result in an apparent barycentre shift detected by our measurements that is smaller by a factor of $1/(1+q)$ and $q/(1+q)$, respectively (Eq. \ref{eq:barycentre}). Therefore we consider host masses of $M_A=0.06\,M_{\odot}$, appropriate for an age of $\sim$3 Gyr \citep{Faherty:2014aa}, and $M_B=q M_A\simeq0.047\,M_{\odot}$ and increased the reference r.m.s. by the respective reciprocal factor, which deteriorates the detection limit. We repeated this with $M_A=0.03\,M_{\odot}$, corresponding to a younger age of $\sim$0.5 Gyr.

\begin{figure} 
\centering
\includegraphics[width=\linewidth]{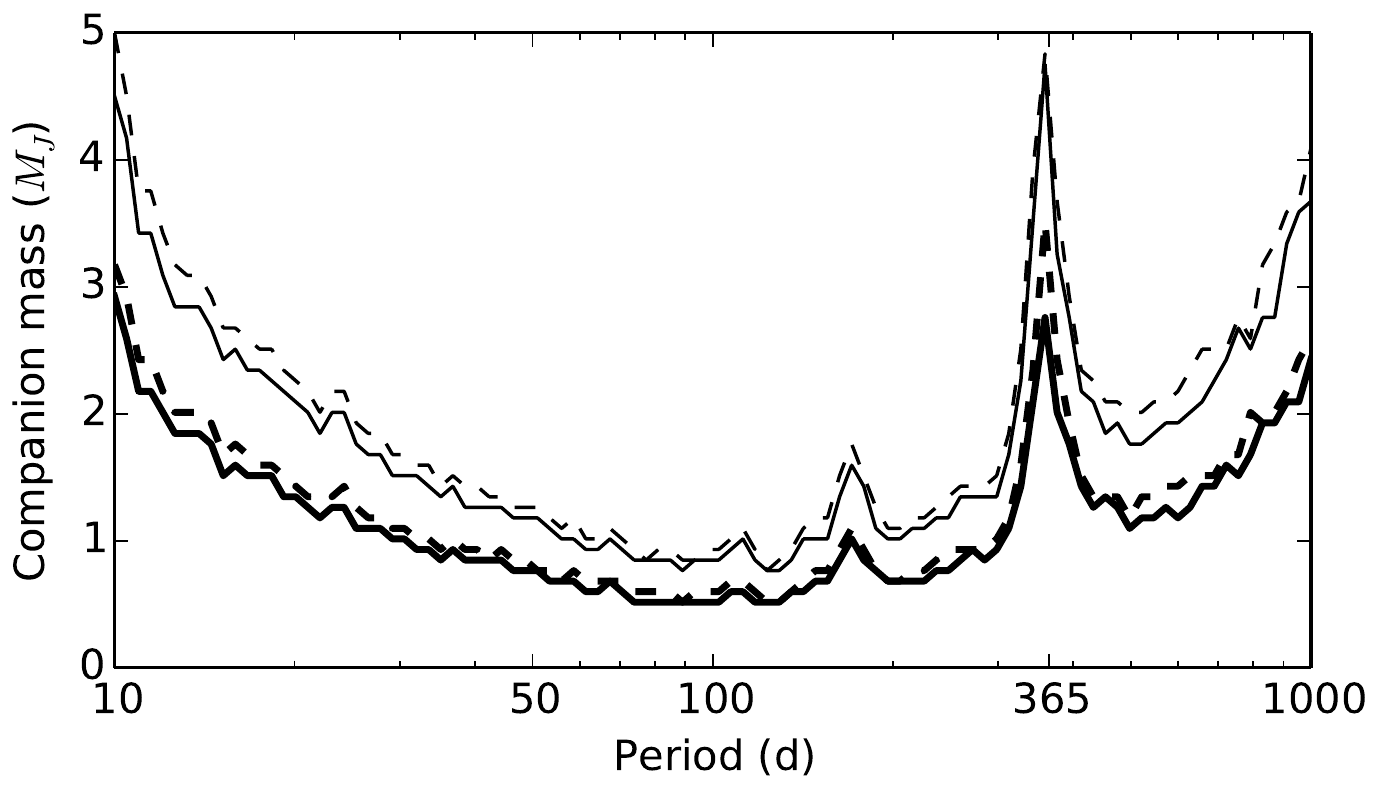}
\caption{Companion mass limits for \dw\ A (solid) and B (dashed) as a function of orbital period. Thick and thin lines correspond to primary masses of $M_A=0.03\,M_{\odot}$ and $0.06\,M_{\odot}$, respectively. The maximum companion mass compatible with the measurements is shown, i.e.\ companions on and above the curve are excluded by the data at 3-$\sigma$ confidence.} 
\label{fig:detLimits}
\end{figure}

Figure \ref{fig:detLimits} shows the maximum mass of any putative companion to \dw\ A and B as a function of its orbital period. The sensitivity deteriorates for very short periods because the astrometric signal decreases, at long periods because those exceed the measurement timespan, and at 365 d because of correlation with the parallax signal. However, for orbital periods between $\sim$20 and $\sim$300 days ($\sim$0.05--0.4 au), the FORS2 data exclude at 3-$\sigma$ confidence the presence of planetary companions with masses larger than $\sim$2 Jupiter masses ($M_\mathrm{J}$) around A and B, assuming a primary mass corresponding a 3 Gyr age. At 100 d period, any companion more massive than Jupiter is excluded. If \dw\ is younger $\sim$0.5 Gyr, the detection limits are a factor of two better. These limits also exclude the presence of a third brown dwarf in the system within the explored period range, except for a potential companion that has equal mass and $I$-band luminosity as the object it orbits, because for these systems the induced photocentre shift would be null.

We thus excluded that either of the \dw\ components harbours a giant planet more massive than twice the mass of Jupiter in a short period (20--300 d) orbit, which disagrees with the conclusions of \citet{Boffin:2014aa}. So far, the system appears to consist of only two components.

\section{Conclusions}
We analysed multi-epoch imaging observations of \dw\ and measured the astrometric motions of both the L- and T-dwarf components that show considerable orbital motion. By modelling the system's barycentre, which follows the standard parallax and proper motion, we determined the mass ratio of the two brown dwarfs to be $q=0.78 ^{+0.10}_{-0.09}$, which is a model-free measurement derived directly from the astrometry and that does neither depend on the parallax value nor on the orbital parameters. In comparison with other ultracool binaries, this mass ratio is low but not exceptional \citep{Liu:2010fk}. We also updated the parallax to $500.51\pm0.11$ mas. 

As one of the closest stellar objects to the Sun, \dw\ is an ideal target for astrometric planet search. We used the FORS2 astrometry to search for a planet in the \dw\ system and found no significant signal. We excluded the presence of Jupiter-mass planets in short period orbits. Continued monitoring with FORS2 or other instruments will improve the sensitivity to lower-mass planets and complementary imaging, spectroscopic, and photometric observations will allow us to further investigate the presence of extrasolar planets in the immediate solar neighbourhood. The results presented here contribute to the detailed characterisation of \dw, {which is a fundamental system for  understanding brown dwarfs and their physics.} Our mass ratio determination helps with the interpretation of spectroscopic and photometric observations, leading to a better picture of the properties of \dw.

\section*{Acknowledgements}
J.S. is supported by an ESA Research Fellowship in Space Science. This research made use of the databases at the Centre de Donn\'ees astronomiques de Strasbourg (\url{http://cds.u-strasbg.fr}); of NASA's Astrophysics Data System Service (\url{http://adsabs.harvard.edu/abstract\_service.html}); of the paper repositories at arXiv;  and of Astropy, a community-developed core Python package for Astronomy \citep{Astropy-Collaboration:2013aa}.  

\bibliographystyle{mnras}
\bibliography{/Users/sahlmann/Dropbox/L16/papers}

\thispagestyle{empty}

\appendix
\begin{table*}
\caption{{Epoch astrometry of \dw A and B in the ICRF, after the effects of DCR have been removed. The quoted uncertainties correspond to the photocentre precision. The conversion to ICRF introduces an additional uncertainty of 80 mas in both RA and Dec, which was not incorporated here.}}
\centering
\begin{tabular}{ccccccccc}
\hline
& \multicolumn{4}{c}{Primary} & \multicolumn{4}{c}{Secondary}\\
Epoch & RA & $\sigma_\mathrm{RA}$ & Dec & $\sigma_\mathrm{Dec}$ &RA & $\sigma_\mathrm{RA}$ & Dec & $\sigma_\mathrm{Dec}$\\
(MJD) & (deg) & (mas)& (deg) & (mas)& (deg) & (mas)& (deg) & (mas)\\
\hline
56396.9941 & 162.31119262 & 0.18 & -53.31833644 & 0.23 & 162.31071214 & 0.23 & -53.31805513 & 0.25 \\
56402.0614 & 162.31116077 & 0.16 & -53.31832531 & 0.18 & 162.31068269 & 0.18 & -53.31804499 & 0.18 \\
56408.0411 & 162.31112537 & 0.18 & -53.31831183 & 0.20 & 162.31064843 & 0.21 & -53.31803272 & 0.21 \\
56418.0342 & 162.31107044 & 0.19 & -53.31828831 & 0.21 & 162.31059589 & 0.21 & -53.31801145 & 0.23 \\
56424.0872 & 162.31103953 & 0.22 & -53.31827345 & 0.25 & 162.31056730 & 0.25 & -53.31799796 & 0.26 \\
56432.9754 & 162.31099898 & 0.18 & -53.31825165 & 0.20 & 162.31052781 & 0.20 & -53.31797790 & 0.21 \\
56438.0317 & 162.31097762 & 0.21 & -53.31823918 & 0.24 & 162.31050839 & 0.23 & -53.31796661 & 0.24 \\
56443.9928 & 162.31095466 & 0.22 & -53.31822436 & 0.25 & 162.31048650 & 0.26 & -53.31795325 & 0.26 \\
56447.9940 & 162.31094034 & 0.18 & -53.31821458 & 0.20 & 162.31047347 & 0.20 & -53.31794448 & 0.21 \\
56452.9718 & 162.31092407 & 0.15 & -53.31820274 & 0.16 & 162.31045826 & 0.16 & -53.31793372 & 0.17 \\
56456.9943 & 162.31091192 & 0.16 & -53.31819350 & 0.18 & 162.31044753 & 0.18 & -53.31792530 & 0.19 \\
56459.9747 & 162.31090367 & 0.17 & -53.31818664 & 0.19 & 162.31043962 & 0.19 & -53.31791918 & 0.20 \\
56465.9674 & 162.31088850 & 0.18 & -53.31817356 & 0.20 & 162.31042610 & 0.21 & -53.31790745 & 0.22 \\
56693.2481 & 162.31035855 & 0.14 & -53.31827471 & 0.16 & 162.30995579 & 0.15 & -53.31806324 & 0.16 \\
56702.1882 & 162.31029478 & 0.12 & -53.31827399 & 0.14 & 162.30989451 & 0.13 & -53.31806508 & 0.14 \\
56726.1436 & 162.31011923 & 0.14 & -53.31825808 & 0.15 & 162.30972604 & 0.15 & -53.31805525 & 0.15 \\
56735.1809 & 162.31005308 & 0.14 & -53.31824667 & 0.16 & 162.30966312 & 0.15 & -53.31804615 & 0.17 \\
56746.1227 & 162.30997580 & 0.14 & -53.31822925 & 0.15 & 162.30958868 & 0.16 & -53.31803161 & 0.16 \\
56757.1145 & 162.30990168 & 0.13 & -53.31820849 & 0.15 & 162.30951851 & 0.15 & -53.31801384 & 0.15 \\
56772.9997 & 162.30980342 & 0.12 & -53.31817378 & 0.14 & 162.30942424 & 0.14 & -53.31798334 & 0.15 \\
56782.9836 & 162.30974806 & 0.15 & -53.31815003 & 0.17 & 162.30937176 & 0.16 & -53.31796206 & 0.16 \\
56795.9827 & 162.30968407 & 0.13 & -53.31811779 & 0.14 & 162.30931272 & 0.14 & -53.31793353 & 0.15 \\
\hline
\end{tabular}
\label{tab:epastr}
\end{table*}

\bsp	
\label{lastpage}
\end{document}